\documentclass[%
 reprint,
 amsmath,amssymb,
 aip,apl
]{revtex4-2}

\usepackage{graphicx}
\usepackage{dcolumn}
\usepackage{bm}
\usepackage{enumerate}
\usepackage{comment}
\usepackage{xcolor}
\usepackage{hyperref}
\usepackage{units}
\usepackage{booktabs}
\usepackage{adjustbox}

\begin{document}
\newcommand{\C}[2]{{#1}_\mathrm{{#2}}}
\newcommand{\red}[1]{\emph{\textcolor{red}{#1}}}
\newcommand{\blue}[1]{\emph{\textcolor{blue}{#1}}}
\preprint{APS/123-QED}

\title{Substrate insulated Josephson junctions for superconducting quantum circuits}

\author{U. Strobel}
\affiliation{Physikalisches Institut, Karlsruher Institut für Technologie}
\author{L. Radtke}
\affiliation{Physikalisches Institut, Karlsruher Institut für Technologie}
\author{L. Kamps}
\affiliation{Physikalisches Institut, Karlsruher Institut für Technologie}
\author{J.N. Voss}
\affiliation{Physikalisches Institut, Karlsruher Institut für Technologie}
\author{J. Lisenfeld}
\affiliation{Physikalisches Institut, Karlsruher Institut für Technologie}
\author{J. Luo-Hofmann}
\affiliation{Fraunhofer-Institut für Elektronische Nanosysteme, Chemnitz}
\author{D. Reuter}
\affiliation{Fraunhofer-Institut für Elektronische Nanosysteme, Chemnitz}
\author{S. Masis}
\affiliation{Physikalisches Institut, Karlsruher Institut für Technologie}
\author{A.V. Ustinov}
\affiliation{Physikalisches Institut, Karlsruher Institut für Technologie}
\affiliation{Institut für Quantenmaterialien und Technologie, Karlsruher Institut für Technologie}
\author{H. Rotzinger}
\email{rotzinger@kit.edu} 
\affiliation{Physikalisches Institut, Karlsruher Institut für Technologie}
\affiliation{Institut für Quantenmaterialien und Technologie, Karlsruher Institut für Technologie}

\date{\today}

\begin{abstract}
We have developed a fabrication technique for Josephson junctions that employs a three-dimensional patterned, low-loss substrate instead of commonly used organic resists.
The technique enables the fabrication of high-quality trilayer junctions from a wide range of geometries and materials, including high-melting-point superconductors such as tantalum or niobium.
The junction electrodes are free from intentionally introduced oxides and organic materials, which are known sources of decoherence.
We fabricate and characterize underdamped Nb/AlO$_x$/Nb junctions of different sizes in several geometries.
Such junctions enable manufacturing of quantum circuits operating at higher speeds and elevated temperatures.
\end{abstract}

\maketitle

Superconducting quantum circuits critically depend on the quality of their Josephson junctions (JJs), most commonly implemented as Al/AlO$_x$/Al trilayers. There are two main reasons for this approach: The AlO$_x$ exceptional quality as a tunnel barrier~\cite{Murray2021}, and the technologically relevant low melting point of aluminum. 
The latter allows a wide range of fabrication possibilities, such as the use of organic resist lift-off masks and angle-dependent patterning~\cite{niemeyer1974,potts2001,dolan1977}.
Despite these advantages, the relatively low superconducting gap energy of aluminum limits the operating temperature $\lesssim\unit[300]{mK}$ and circuit frequencies $\lesssim\unit[40]{GHz}$. 

Josephson junctions with a higher superconducting gap can overcome these limitations and allow operation of e.g. SQUIDs or qubits at much higher temperatures~\cite{clarke2006, anferov2024a} and frequencies. 
Unfortunately, many well-established thin-film superconductors with an energy gap higher than Al (e.g., Ta, TiN, Nb, NbTi) also have significantly higher melting points and therefore require higher deposition temperatures when the materials are thermally evaporated. Protecting organic resists from burning becomes challenging, and the junction quality is reduced~\cite{martinis2005,martinis2009}. Although other deposition techniques, such as magnetron sputter deposition, can produce high-quality Nb/AlO$_x$/Nb JJs\cite{clarke2006}, this technique requires the addition of a lossy dielectric in the immediate vicinity of the JJs, which introduces excess noise~\cite{bhushan1991,tolpygo2010,meckbach2013,kempf2013,gronberg2017,Dutta2004,Yu2005,Lisenfeld2007,Weides2011,Li2026}. Recent developments have eliminated the presence of these dielectrics through a wet etching post-processing step~\cite{anferov2024a} or via a flip chip technique~\cite{Wang2026}. 

To avoid aggressive oxides etching near the delicate junction, we propose~\cite{rotzinger2025} an alternative approach to junction-electrode isolation based on profiling the substrate material. A step profile electrically disconnects the metal films deposited below and above the step. 
We use this geometric feature to shape and insulate the JJ barrier and electrodes. No additional lossy dielectric material is deposited, nor do organic resist residues contaminate the junction vicinity, because our method allows thorough acid cleaning of the wafer immediately before deposition of the junction metals. The only dielectric seen by the JJ is the low-loss substrate crystal, typically a high resistivity Si or sapphire~\cite{martinis2014}.  
We envisage Josephson junctions that are largely independent of the superconductor, tunneling barrier, deposition techniques, or cleaning methods used. We discuss several implementation possibilities below. 

\begin{figure*}[t]
  \centering
  \includegraphics[width=1\textwidth]{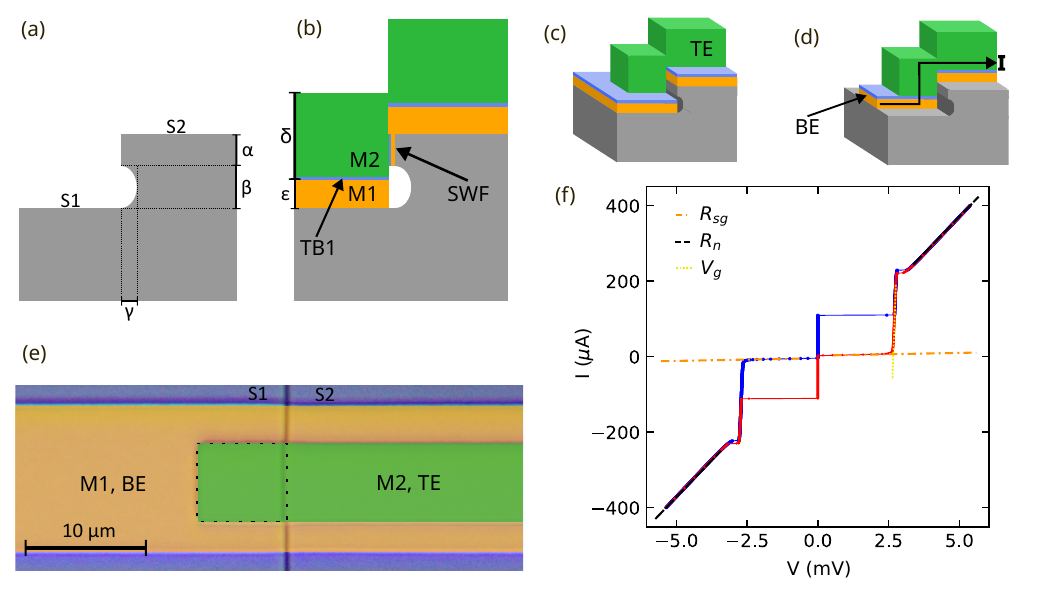}
  \caption{Step edge insulated tunnel junction. $\textbf{(a)}$ Substrate cross-section after forming the step with an undercut. $\textbf{(b)}$ Trilayer deposition. 
  To ensure M2 continuity across the step, we require $\delta > \alpha + \beta -\epsilon$. $\textbf{(c)}$ The top electrode (TE) is formed by lithography. $\textbf{(d)}$ The bottom electrode (BE) is formed by lithography. The current direction is shown by the black arrow. $\textbf{(e)}$ False-color optical image of an Nb/AlO$_x$/Nb SEI junction. The junction area is enclosed by the dashed-line rectangle. $\textbf{(f)}$ A typical I-V characterization of a junction with nominal area $A=\unit[100]{\mu m^2}$ at $4.2\,\text{K}$ indicates an underdamped junction with a linear dissipative branch for bias current above $I_\mathrm{c}=\unit[110]{\mu A}$. The return path shows a pronounced hysteresis with a low retrapping current $I_\mathrm{r}$. A gap voltage of $V_\mathrm{g} = 2 \Delta/e = \unit[2.78]{mV}$ was extracted at the interpolation of the yellow dotted line with the x-axis~\cite{golubov1995proximity}. The line is a linear fit between the 10-th and 50-th percentile of current in the $[2,3.5]\,\mathrm{mV}$ window. Normal (subgap) resistance of $R_\mathrm{n}=\unit[12.7\pm0.1]{\Omega}$ ($R_\mathrm{sg}\approx\unit[700]{\Omega}$) was extracted by fitting the black (orange) dashed (dashed dotted) line above $\unit[3.5]{mV}$ (between $\unit[0.9]{mV}$ and $\unit[2.2]{mV}$) and measuring the inverse slope ($I(\unit[2]{mV})/\unit[2]{mV}$). The knee adjacent to the normal resistance branch is associated with the proximity effect of Nb on the not fully oxidized Al~\cite{golubov1995proximity}.
  }
  \label{fig:SEI_combo}
\end{figure*}

\textit{Step edge insulated} (SEI) junction.
The step profile consists of an overhang~\footnote{The overhang must not necessarily be vertical, as long as it is high enough to allow for an undercut to form.} of height $\alpha$ and an undercut of height $\beta$ and depth $\gamma$ (Fig.~\ref{fig:SEI_combo}~(a)).   
In the following stage, the M1/TB1/M2 trilayer is deposited in situ on S1 and S2 simultaneously, where M1 (M2) is the top (bottom) junction metal of thickness $\delta$ ($\epsilon$), TB1 is the tunnel barrier of a negligible thickness, and S1 (S2) is the lower (upper) floor of the patterned substrate.
The overhang sidewall is generally also coated with M1, creating a thin parasitic sidewall film (SWF) that may short the junction. The effect is less pronounced with anisotropic deposition methods, such as e-beam evaporation or laser ablation, applied at an appropriate deposition angle; however, it is still non-negligible.
The purpose of the undercut in the substrate step is to prevent this short by setting $\epsilon<\beta$ and $\gamma >0.5 \epsilon$. 
TB1 can be formed by deposition of a dielectric or, often, by oxidation of the bottom electrode (BE). M2 can generally be formed by a different technique than M1, e.g., isotropic atomic layer deposition (ALD) or sputtering, provided the vacuum is not broken, to ensure clean interfaces across the junction. If M2 is deposited with an anisotropic method, we find it feasible depositing at an angle facing the overhang to avoid growth boundaries between the bottom and top M2.

The bottom electrode (BE) and the top electrode (TE) are located on S1 and S2, respectively, and are laterally defined by two consecutive lithography steps with organic or hard masks and dry or wet etching. The junction size is determined by the TE lithography resolution and alignment precision (Fig.~\ref{fig:SEI_combo}). 
 
\begin{figure*}[t]
\includegraphics{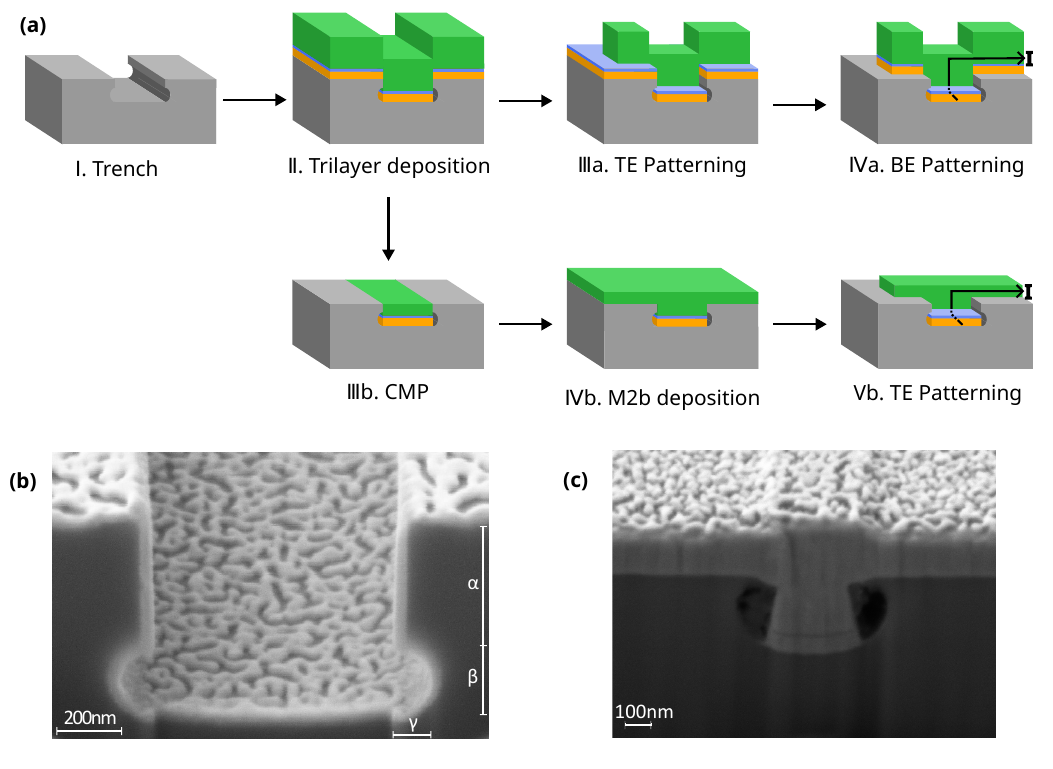}
\caption{\label{fig:cSEI_Implementation}$\textbf{(a)}$~cSEI (steps I-II-IIIa-IVa) and p-cSEI (steps I-II-IIIb-IVb-Vb) junctions.
$\textbf{(b)}$~Cross-sectional scanning electron microscope (SEM) image of a trench, taken at $55^\circ$ tilt after a $\unit[4]{nm}$ Au sputter and a focused ion beam (FIB) cut. The $\unit[800]{nm}$ wide trench is etched into a silicon substrate using the Bosch undercut process. The undercut dimensions are in good agreement with the design parameters of $\alpha = 450\,$nm, $\beta = 250\,$nm and $\gamma = 100\,$nm. 
$\textbf{(c)}$~p-cSEI Nb/AlO$_x$/Nb/Nb junction cross-sectional SEM image. 
}
\end{figure*}
The \emph{cross step edge insulated} (cSEI) junction removes the SEI junction alignment accuracy limitation by using a second step edge, parallel to the first step, to define a BE confining trench. TE is lithographically patterned in a direction perpendicular to the trench, with a large alignment margin. The junction lateral dimension is defined by the intersection of the trench and the TE line, and is only limited by the lithography resolution. The fabrication sequence is similar to the SEI junction (Fig.~\ref{fig:cSEI_Implementation}). 
In the case of consecutive layer deposition, M2 fully protects M1 and TB1 from potential shorts and aggressive etching or cleaning steps.
Deposition onto S1 at the bottom of a narrow, deep trench requires a deposition angle close to normal incidence, potentially causing enhanced SWF formation and irregularities at the boundary where M2 grown on S1 meets M2 grown on S2. 

A planarized variation of the cSEI junction (\emph{p-cSEI} junction) resolves the latter problem by adding a planarization step~\cite{bao1995,tolpygo2015} (e.g., chemical-mechanical polishing (CMP)) after the trilayer deposition, followed by ion mill cleaning and an additional M2b deposition (Fig.~\ref{fig:cSEI_Implementation}(a), steps I-II-IIIb-IVb-Vb). 
After planarization, subsequent steps, e.g., to form interconnects, are simpler because there are no large height differences to consider. 

\begin{figure}[t]
\includegraphics{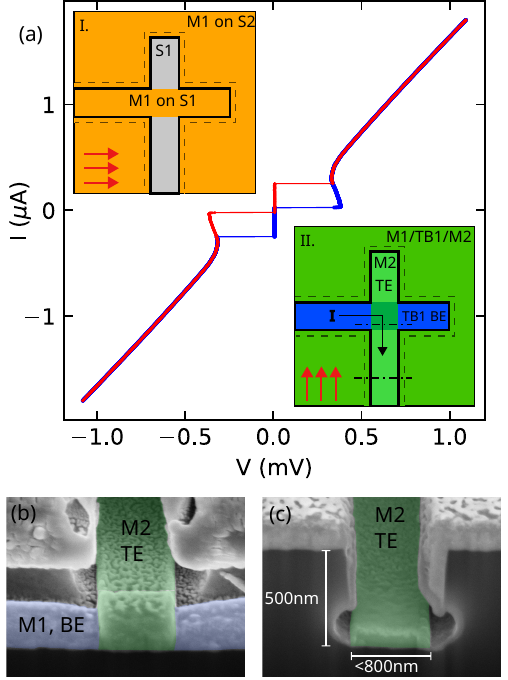}
\caption{\label{fig:MT_Implementation}  
$\textbf{(a)}$~I-V characterization of an Al/AlO$_x$/Al planarized Manhattan Trench junction with area $A=0.25\mu\text{m}^2$ at $15\,\text{mK}$.
Inset $\textbf{I.}$~Top-view schematic after etching two perpendicular trenches in the substrate and depositing M1 along the red arrows. The cross shape of the trenches' top (in the S2 plane) is depicted with a solid black line. Around it, the dashed grey line depicts the undercut (in the S1 plane). The walls of the north--south trench cast a shadow on the S1 floor of the trench and protect it from deposition.
Inset $\textbf{II.}$~TB1 deposition followed by M2 deposition along the red arrows. Now, the east--west trench floor is in the shade. The junction is formed in the overlap of the two trenches. The black arrow illustrates the electrical current direction. 
$\textbf{(b)}(\textbf{(c)})$~False-color SEM image of the FIB-cut Nb/AlO$_x$/Nb cross-sectional view before planarization. The location is indicated by the top (bottom) dash-dotted line in (a)II. 
}
\end{figure}

The \emph{Manhattan Trench} (MT) junction~\cite{potts2001} is formed entirely on S1 at the intersection of two perpendicular trenches. First, M1 is deposited at a steep angle ($30-60^{\circ}$), which exposes S1 of the first trench (T1) but keeps S1 of the second trench (T2) in the shade of the T2 steps, forming the BE in T1. Before M2 deposition, the sample azimuth is rotated by $90^\circ$ so that S1 of the second trench is exposed while the BE is shaded by the T1 steps; the TE is formed in T2. The junction does not touch any walls on which SWF could form ($\delta + \epsilon<\beta$, Fig.~\ref{fig:MT_Implementation}). The trenches can be deeper than in the SEI and the cSEI junctions because the TE does not climb from S1 to S2. An optional planarization step removes the excess trilayer from S2. Outside the trenches, the trilayer on S1 acts as an effective single superconducting layer due to unavoidable pinholes in TB1 that electrically short M1 and M2. As a result, typical single-layer microwave and DC readout and control circuitry can be defined by the same lithography step that formed the junction trenches. In a generalized junction geometry, the S1/M1/TB1/M2 patch is wider than the BE and TE trenches that lead to it, allowing arbitrarily large junctions~\cite{zhang2017}.

We propose an additional closely related junction geometry, the \emph{bridge trench}, where a single trench is interrupted by a short break. The undercuts of the two segments join, creating an S1 tunnel under an S2 bridge, similar to the Dolan-bridge~\cite{dolan1977}. Adjusting the bridge shadow between M1 and M2 depositions allows the formation of a junction on S1, while all material on S2 is sacrificial and may be removed in an optional planarization step. The effective junction dimensions depend on the deposition angles.

\begin{figure}[t]
\includegraphics{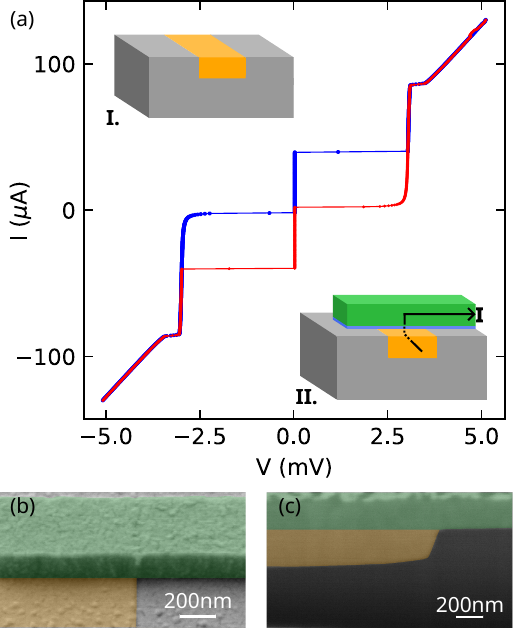}
\caption{\label{fig:tC_combo}
$\textbf{(a)}$~I--V characteristic of an Nb cT junction with $A=16\mu\text{m}^2$ at $50\,\text{mK}$. 
Inset $\textbf{I.}$~Sample state after (1)~no-undercut trench etch ($\gamma\le0$), (2)~M1 deposition ($\epsilon > \alpha+\beta$), and (3)~planarization. BE is formed in the trench. 
Inset $\textbf{II}$~Sample state after (4)~hydrofluric acid dip and ion-mill cleaning to remove the native oxide, (5)~TB1 and M2 deposition, (6)~TE patterning, and (7)~TB1 patterning (optional, possibly with the same mask as the previous step).$\textbf{(b)}$~False-color SEM image of Nb cT junction$\textbf{(c)}$~Cross-sectional SEM image}
\end{figure}

In the \emph{cross-trench} (cT) junction, the wafer is planarized down below S2 after M1 deposition (Fig.~\ref{fig:tC_combo}). Exposing M1 to an aggressive treatment stands in stark contrast to the other trilayer deposition methods described above and may affect the junction quality. Yet, the simplicity of trench formation without an undercut and of all subsequent fabrication steps may outweigh this disadvantage. The junction size is limited only by lithography resolution. 
Unlike the traditional cross-junctions~ \cite{braumuller2015,wu2017}, the BE trench walls leave only the top BE surface exposed, allowing non-isotropic TB1 and M2 deposition. M1 and M2 can be conveniently deposited in separate tools.

Quantum circuits require low-microwave-loss wafers, typically sapphire or high-resistance silicon~\cite{martinis2014}. The step-with-undercut profile required for our junctions could possibly be realized in the former using wet-etch~\cite{shen2011}. We instead use the Bosch process on silicon~\cite{laermer1990,laerme1999} to pattern the latter with high precision and reproducibility (Fig.~\ref{fig:cSEI_Implementation}~(b)). 

We measured the surface roughness of S1 and S2 using an atomic force microscope. For a $250$\,nm-deep trench, the average S1 surface roughness is $S_a \approx 1.2\,$nm, compared with $S_a \approx 0.8\,$nm for S2. Although higher, this value is still acceptable for typical thin films of $20$--$\unit[200]{nm}$ thickness. Further surface treatment, e.g., wet chemical etching, may reduce the roughness if required. To determine the influence of Si surface roughness on the superconducting film quality, we deposited $75\,$nm Nb in a home-built DC magnetron sputtering system with a base pressure of $\sim 1\times10^{-8}$\,mbar. 
Cryogenic measurements showed a critical temperature $T_c = 9.0$\,K for films on S1, similar to the $T_c$ measured for films deposited on pristine substrates.

\begin{figure}[t]
\includegraphics{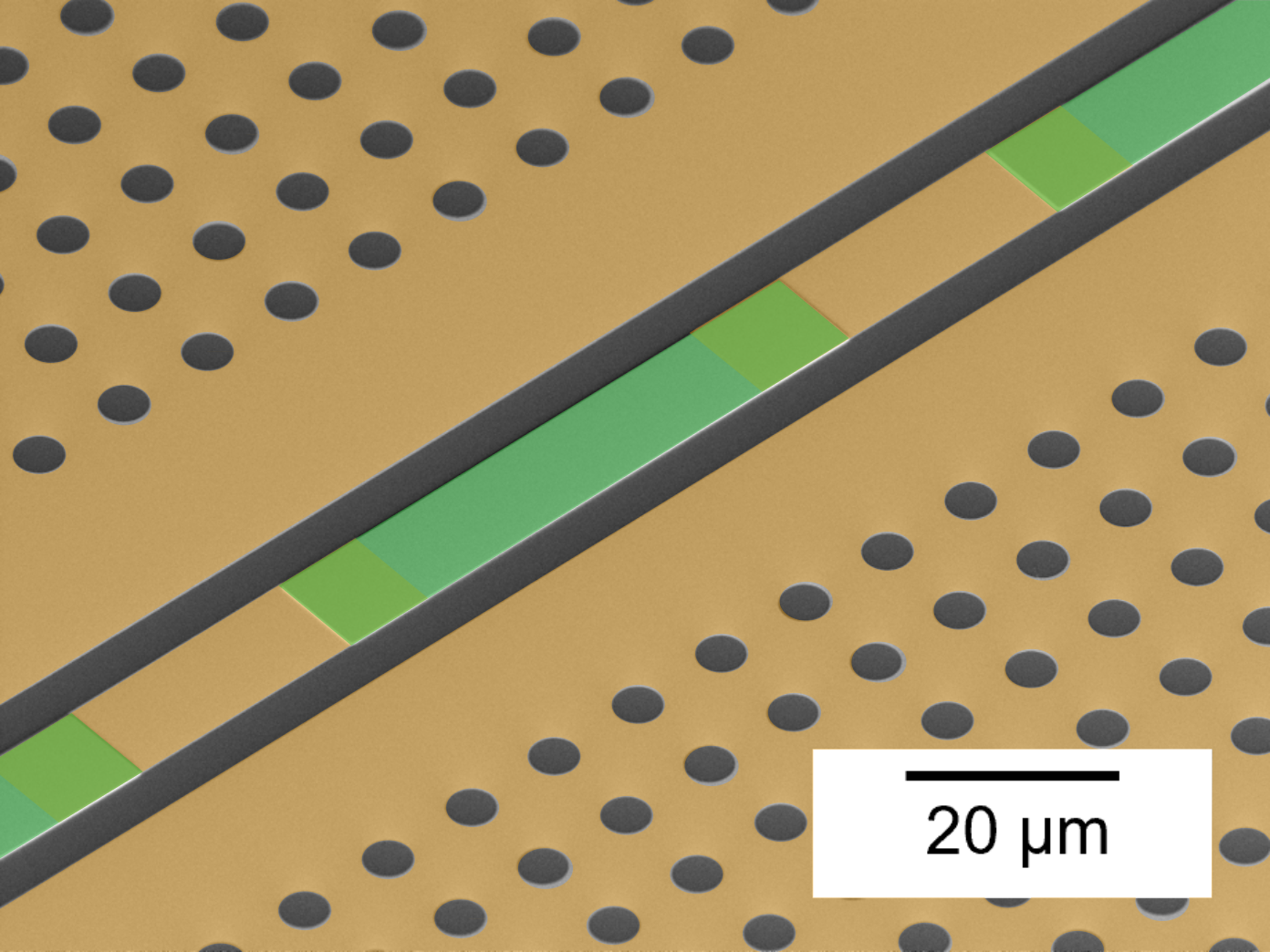}
\caption{\label{fig:ENAS_Resonator}
False-color SEM image of a two-layer niobium CPW resonator structure. Niobium embedded in the substrate is shown in orange, and the second niobium layer in green.}
\end{figure}

To investigate microwave properties of superconducting quantum circuits realized with the CMP technique, we measured niobium $\lambda /4$ coplanar waveguide (CPW) resonators at millikelvin temperatures. The measured resonance frequency was approximately 7\,GHz and the internal quality factor exceeded $3\times10^5$. The most likely sources of loss are the native niobium oxide layer of M1 and the roughness of the argon-milled S1. To test the M1--M2 interface, we introduce a gap in the BE trench, fill both segments with M1, planarize, and bridge the gap with M2 (Fig.~\ref{fig:ENAS_Resonator}). The surface of the first niobium film is ion milled in situ before the second film is deposited. The two-layer resonators show no degradation in quality factor compared with the single-layer resonators.   

\begin{figure}[!t]
\includegraphics[width=\columnwidth]{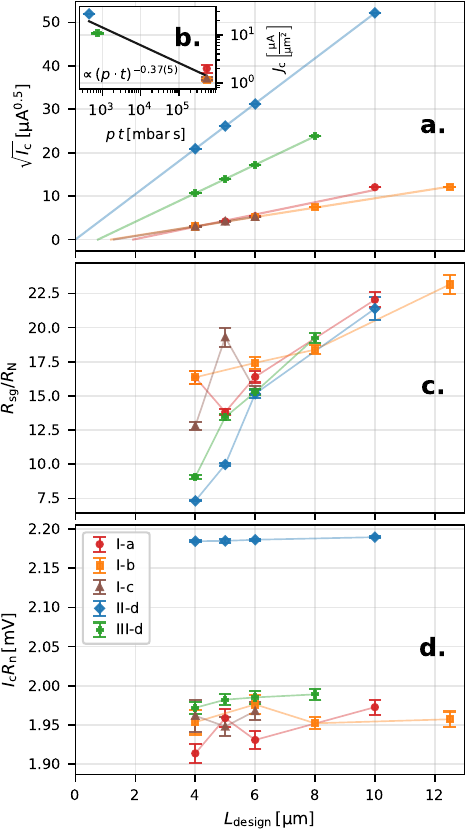}
\caption{\label{fig:Ic_Rsg_IcRn} (a) 
Roman numerals I--III indicate different wafers, while letters a--d indicate the test-chip position on the wafer. $\C{I}{c}$ modeled by Eq.~\ref{eq:Ic} using the fit parameters $u$ and $\C{J}{c}$.
For readability, the vertical axis shows $\sqrt{\C{I}{sw}}$ (instead of $\C{I}{sw}$): The square-root scaling compresses the dynamic range so large-current junctions do not dominate the plot. 
Chip II-d, with the highest $\C{J}{c}$, presented a self heating signature - negative slope near $\C{V}{g}$. For this chip only the $\C{V}{g}$ was taken as the maximum voltage in the $[0,100]\,\mathrm{\mu A}$ window.
(b) The dependence of the critical current density on the product of oxygen partial pressure $p$ and oxidation time $t$~\cite{kleinsasser1995} is fitted by a power law.
(c) The leakage ratio $\C{R}{sg}/\C{R}{n}$ is improved with junction size.
(d) The characteristic voltage $\C{I}{c}\C{R}{n}$. Values are on par with other Nb/AlO$_x$/Nb junctions~\cite{anferov2024a,kempf2013,kaiser2010,Lehnert_1994}.
}
\end{figure}

A set of SEI junctions of various sizes with different oxidation parameters but a common 3D geometry $(\alpha,\beta,\gamma)=\unit[(100,150,50)]{nm}$ 
were characterized at room and cryogenic temperatures (data taken at $\unit[4.2]{K}$ for chips from wafer I and at $\sim\unit[2.5]{K}$ for chips from wafer II and III, as depicted in Fig.~\ref{fig:SEI_combo}~(f)). The junctions are nominally square with an area of $\C{L}{design}^2$. The critical current was extracted with the help of Ambegaokar-Baratoff formula~\cite{tinkham1996} $\C{I}{c}=\C{R}{n}^{-1}(\pi\C{V}{g}/4)\tanh(\C{V}{g}e/4\C{k}{B}T)$. 
We model the critical current density $\C{J}{c}$ of a junction as 
\begin{equation} \label{eq:Ic}
	\C{J}{c}=\C{I}{c}/\C{L}{eff}^2,
\end{equation}
where  $\C{L}{eff}=\C{L}{design}-2u$ is the effective length and $u$ is the undercut parameter. A linear fit of $\sqrt{\C{I}{c}}(\C{L}{design})$ hence allows to identify $\sqrt{\C{J}{c}}$ as the inclination of the line, and $2u$ as the the interception with the x-axis (Fig.~\ref{fig:Ic_Rsg_IcRn}). 

Two size-independent junction figures of merit are the characteristic voltage  $\C{V}{c}\equiv\C{I}{c}\C{R}{n}$ and the leakage ratio $\C{R}{sg}/\C{R}{n}$. High $\C{V}{c}$ indicates that the junction approaches the Ambegaokar–Baratoff tunneling limit, whereas barrier defects/excess conductance typically reduce it.
Likewise, large $\C{R}{sg}/\C{R}{n}$ implies suppressed subgap quasiparticle transport (lower intrinsic dissipation)~\cite{Tolpygo2013}.   
In our case $\C{R}{sg}/\C{R}{n}$ increased with junction size, suggesting that junction edges primarily contribute to the leakage.

In conclusion, we have established a substrate-profiling technique to fabricate oxide-free and resist-residue-free underdamped Josephson junctions of various sizes and geometries. The current--voltage characteristics show pronounced hysteresis, indicating a high junction quality. The technique enables fabrication of quantum circuits across a wide range of materials and deposition conditions.  

Supplementary material presents fabrication details and evaluation of the junction critical currents.

We thank Silvia Diewald and Patrice Brenner for the help with lithography and FIB imaging. This work was supported by funding from the European Research Council (ERC) under the European Union's Horizon 2020 research and innovation program (project {\em Milli-Q}, grant agreement number 101054327). The research was also supported by the Fraunhofer QNC.space project CohereMP and the German Federal Ministry of Education and Research under the Research Program Quantum Systems, through the projects GeQCoS (FZK13N15691), qBriqs
(FZK13N15950) and Qrious (FZK13N17125).
\bibliographystyle{apsrev4-2} 
\bibliography{Substrate_Insulated_JJ_Paper}

\appendix
\title{Supplementary Material for: High Quality Josephson junctions featuring substrate electrode insulation}

\section{Fabrication Process}
\subsection{SEI Josephson junction}
After optical or e-beam resist patterning, a smooth vertical step of depth $\alpha$ is etched anisotropically for 1 minute using an HBr plasma in an inductively coupled plasma (ICP)\footnote{Oxford Plasmalab 100} tool. The HBr flow is $15\,$sccm, the pressure is $3\,$mTorr, the ICP power is $250\,$W, and the rf power is $20\,$W.
We then perform the Bosch deep-silicon etching process.
\begin{enumerate}
    \item [a.] Isotropic passivation in a reactive ion etcher (RIE)\footnote{Sentech SI 220} by applying a $C_4F_8$ plasma for $20\,$s with a gas flow of $25\,$sccm, a pressure of $8\,$Pa and a power of $100\,$W.
    \item [b.] Isotropic etching to a depth $\beta$ using a fluorine--argon plasma ($SF_6$ and Ar) for $18\,$s in the same RIE. The gas pressure is $10\,$Pa, the power is $100\,$W, and the $SF_6$ and Ar gas flows are $15\,$sccm and $60\,$sccm, respectively. The undercut geometry (width and depth) can be fine-tuned by adjusting the pressure, gas flows, and plasma power.
\end{enumerate}
After stripping the resist and etch residues~\footnote{Techniclean CA25} the chip clean is followed by a piranha solution step. Prior to metal deposition, we remove the native silicon oxide using hydrofluoric acid (HF).

To form the SEI junction, the same sputtering system is used to deposit the $50/8/200\,\mathrm{nm}$ Nb/Al/Nb trilayer at a normal angle to the wafer. TB1 is formed by in situ oxidation the Al layer. The top Nb layer is etched using an $SF_6$/Ar plasma in the RIE, with the AlO$_x$ layer acting as an etch-stop layer. Before the second lithography step, we remove the AlO$_x$ layer and the bottom Nb using a chlorine-based plasma in the ICP.
The ICP settings are: Cl/Ar flow of 2/12 sccm, pressure of 10 mTorr, ICP power of 100 W, and RF power of 100 W.

\subsection{p-cSEI Josephson junction}

For the p-cSEI junctions, the wafer etching is performed similarly to the SEI junctions, except that the etching times of the HBr and SF6/Ar etch are increased to 180\,s and 24\,s, respectively, resulting in $(\alpha,\beta,\gamma)=(300,250,100)\,\mathrm{nm}$.
A 75/7/2/250\,nm Nb/AlO$_x$/Al/Nb stack is deposited at a normal angle to the wafer using an e-beam shadow evaporation tool\footnote{Plassys Model MEB 550 S} at a chamber pressure of $\sim 2\times10^{-8}$\,mbar. The oxide layer is formed in situ by controlled oxidation of the deposited Al.
The wafer is then CMP polished \footnote{Bruker TriboLab CMP system with a 15\,nm particle-size slurry (CS30-316P)}, at a flow rate of $\unit[33]{ml/min}$ and a force of $\unit[100]{N}$, until Si is exposed on S2.
The wafer is cleaned and returned to the evaporation tool for ion milling, followed by deposition of $\unit[150]{nm}$ Nb.
The e-beam lithography and patterning steps are similar to the SEI implementation.

\subsection{MT Josephson junction}
Wafer etching of the MT Josephson junction is similar to the p-cSEI junction, but with the first HBr anisotropic etch time increased to five minutes, resulting in $(\alpha,\beta,\gamma)=(450,250,100)\,\mathrm{nm}$, sufficient for the Manhattan geometry.
Consequently, 60/80\,nm Al/AlO$_x$/Al or 75/7/1/3/75\,nm Nb/Al/Al/AlO$_x$/Al/Nb are deposited using the evaporation tool. The deposition angle in both implementations is set to $60^\circ$ and crucible to wafer distance is above $30\,\mathrm{cm}$. The wafer is rotated azimuthally by $90^\circ$ before depositing the TE.
In the niobium implementation, the second Al film is deposited after the $90^\circ$ rotation to ensure aluminum-oxide capping of the BE sidewall.
The CMP is performed similarly to the p-cSEI junctions, with the only modification being a reduced force of 50 N for the all-aluminum implementation.

\subsection{cT Josephson junction}
For the cT junction, an 8-inch silicon wafer is patterned using an O$_2$/SF$_6$ plasma (20/100 sccm) at $\unit[600]{W}$ and $\unit[120]{mTorr}$ for 28 s, resulting in a $(\alpha,\beta,\gamma)=(400,0,0)\,\mathrm{nm}$ trench. The wafer is then cleaned in a DMSO bath at $60\,^\circ\mathrm{C}$ for 20 min, followed by an EKC bath at $60\,^\circ\mathrm{C}$ for 20 min. Next, SiO$_x$ is removed in situ by ion milling, and a $\unit[600]{nm}$ Nb film is deposited by DC magnetron sputtering at $\unit[0.3]{nm/s}$ with $\unit[600]{W}$ power and an initial chamber pressure below $\unit[2\times 10^{-8}]{mbar}$. Planarization is performed\footnote{LK394C3 slurry} and optimized to ensure dishing of less than $\unit[10]{nm}$ across the wafer. The surface roughness of the niobium and silicon after CMP was measured by AFM, yielding $S_a \approx 0.2\,$nm.
The wafer is diced into $20 \times 20~\unit{mm^2}$ chips and subsequently cleaned using piranha solution and HF. After ion-mill cleaning, a 2/3/150 nm AlO$_x$/Al/Nb stack is deposited in our evaporation tool. The first Al thickness is chosen such that no continuous metallic Al film remains after oxidation, which could otherwise shorten the junction. Junction patterning is similar to the p-cSEI process.

\section{Analysis}
An alternative procedure to determine $\C{I}{c}$ is by a per-chip fit of the switching current $\C{I}{sw}$.

The discrepancy between the measured $\C{I}{sw}$ and the underlying $\C{I}{c}$ surpasses $100\%$ for the smaller junctions of high oxidation. Sub-micron junction size is of particular interest to our intended application of mm-wave frequency qubits. An accurate fit that takes this correction into account reduces the estimated value of $u$ by about $\unit[1]{\mu m}$ as compared to the simple $\C{I}{c}\approx\C{I}{sw}$ approximation.
 
\begin{figure}[!t]
\includegraphics[width=\columnwidth]{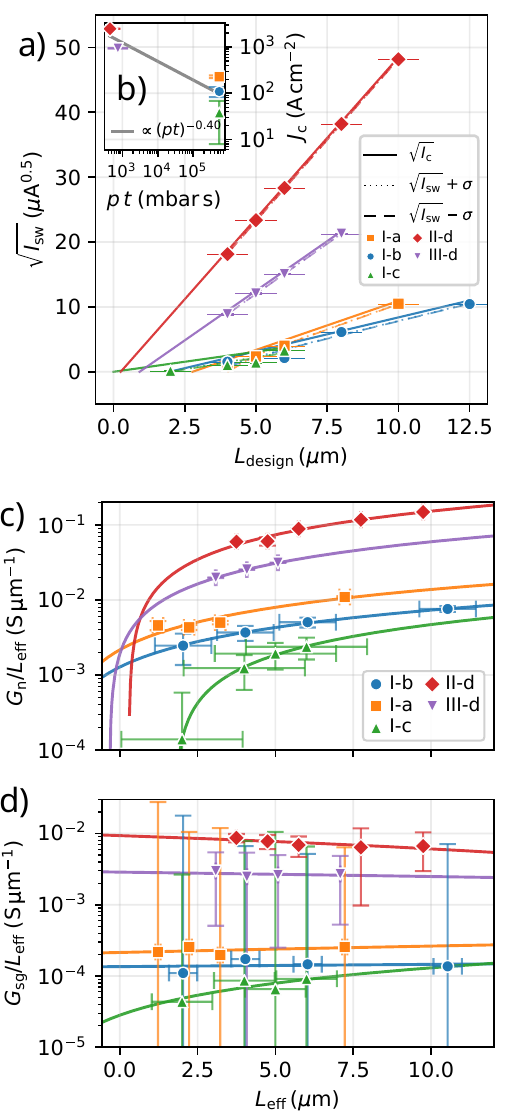}
\caption{\label{fig:Ic} (a) 
Roman numerals I--III indicate different wafers, while letters a--e indicate the test-chip position on the wafer. $\C{I}{sw}$ is fitted per chip by Eq.~\ref{eq:Isw}, with $\C{I}{c}$ modeled by Eq.~\ref{eq:Ic} using the fit parameters $u$ and $\C{J}{c}$. For readability, the vertical axis shows $\sqrt{\C{I}{sw}}$ (instead of $\C{I}{sw}$): The square-root scaling compresses the dynamic range so large-current junctions do not dominate the plot, and it linearizes the quadratic dependence in Eq.~\ref{eq:Ic}. In this representation, the parabolic fit becomes a straight line; its x-intercept equals $2u$, and its slope equals $\sqrt{\C{J}{c}}$.
(b) The dependence of the critical current density on the product of oxygen partial pressure $p$ and oxidation time $t$~\cite{kleinsasser1995} is fitted by a power law.
(c)[(d)])Normal [subgap] conductance $G_\mathrm{n[sg]}$. A linear fit identifies the surface (edge) conductance as the slope (y-intercept). A logarithmic scale is used for clarity.  
}
\end{figure}

The switching current of an underdamped junction is approximated    by~\cite{garg1995}
\begin{equation} \label{eq:Isw}
	\langle I_\mathrm{sw}\rangle \approx
	I_\mathrm{c}\left[
	1-
	\left(\frac{k_\mathrm{B}\Theta_\mathrm{eff}}{U_\mathrm{c}}\ln X\right)^{2/3}
	\right],	
\end{equation}
where 
$X \;=\; \frac{12}{5\pi}\,
\frac{I_\mathrm{c}}{\dot I\,R_\mathrm{eff}C}\,
\left(\frac{k_\mathrm{B}\Theta_\mathrm{eff}}{U_\mathrm{c}}\right)^{2/3}
,$ and the critical barrier height in the cubic-potential approximation is given by $U_\mathrm{c}=\frac{4\sqrt{2}}{3}\,E_\mathrm{J}$. We approximate the effective noise temperature $\C{\Theta}{eff}\approx\unit[7]{K}$ based on a measurement of one of the junctions in a setup with similar cryogenic RC filters~\cite{kammerer2025}. The estimated capacitance~\cite{xiong2018} is $C=\C{L}{eff}^2\C{C}{0}/\ln(\C{J}{0}/\C{J}{c})$, with $\C{C}{0}=\unit[419]{fF/\mu m^2}$ and $\C{J}{0}=\unit[783.5]{\mu A/\mu m^2}$. This estimate is suitable for our range of $\C{J}{c}$, and yielded a $22\%$ deviation from our reference junction~\cite{kammerer2025}, where the capacitance was derived from a measured plasma frequency of $\C{\omega}{p}/2\pi = \unit[138]{GHz}$. We approximate the effective junction resistance at the escape attempt frequency as $R_\mathrm{eff}=Q/\omega_\mathrm{p0}C$, where $\omega_\mathrm{p0}=\sqrt{2\pi I_\mathrm{c}/\Phi_\mathrm{0}C}$ is the plasma frequency, $Q=4I_\mathrm{c}/\pi I_\mathrm{r0}$ is the quality factor~\cite{tinkham1996}, and $I_\mathrm{r0}$ is the retrapping current of the junction at zero temperature, approximated by the measured retrapping current. The bias-current ramp rate $\dot{I}\propto X^{-1}$ is approximated by the bias-current step size divided by the dwell time of $\unit[20]{ms}$. The intrinsic spread $\C{I}{sw}$ is also predicted analytically~\cite{garg1995}, but it is much smaller than the spread between different junctions, which we attribute to fabrication defects and measurement-setup inaccuracies (see the overlapping dashed and dotted lines in Fig.~\ref{fig:Ic}~(a)).

The fit results are depicted in Fig.~\ref{fig:Ic}~(a), and summarized in Table~\ref{tab:Summary}. Data for junctions from the same chip are follow a single straight line; however, the three chips from wafer I have different $\C{J}{c}$ and $u$ despite the simultaneous fabrication. The differences are attributed to the different deposition angles experienced by the different locations on the wafer, causing the substrate step to cast a different shadow.   

The relatively large fitted undercut values of the undercut parameter $u$, beyond the expected lithography misalignment, also suggest irregularities at the edges. To gain some insight, we model the junction normal and subgap conductance values to have two contributions each, one coming from the barrier film between the two superconductors, and the other is related to inhomogeneity effects at the junction edge. While the first scales with the area, the second scales with the perimeter of the junction:
$$
\C{G}{i}=\frac{1}{\C{R}{i}}=\frac{\C{L}{eff}^2}{\C{R}{i,A}}+\frac{\C{L}{eff}}{\C{R}{i,L}},
$$
where $\mathrm{i}\in(\mathrm{n,sg})$ and A [L] subscripts designate the normal (subgap) area-specific [length-specific] resistance. Dividing both sides by $\C{L}{eff}$ allows us to perform linear fits to extract the resistance parameters (Fig.~\ref{fig:Ic}, Table \ref{tab:Summary}).   

Using the fit parameters from Fig.~\ref{fig:Ic}(a,c), we obtain $\C{J}{c}\C{R}{n,A}\approx\unit[1.6]{mV}$, similar to typical niobium Josephson junctions~\cite{anferov2024a,kempf2013, kaiser2010}, and indeed measure comparable $\C{I}{sw}\C{R}{n}$ values in our largest junctions. 
Due to the steep sub-gap fit angle and insufficient number of data pointG, the $\C{R}{n,L}$ cannot be extracted with sufficient accuracy. For the subgap conductance, the fit results in almost horizontal lines that provide an accurate measure for $\C{R}{sg,L}$ but not for $\C{R}{sg,A}$. This result suggests the junctions' quality can be further improved by suppressing the edge conductance, e.g., by an isotropic etching step.  
The $\C{R}{sg}/\C{R}{n}$ ratio demonstrated a trend of  increase with junction size, that was only limited by the maximum area we attempted to $\C{R}{sg}/\C{R}{n}\approx50$.

\begin{table}[t]
\caption{\label{tab:Summary}Summary of fitted parameters}
\centering
\begin{tabular}{lccccc}
  \hline
  Chip & $J_c$ & $u$ & $R_\mathrm{A,n}$ & $R_\mathrm{L,sg}$ \\
   & $\mathrm{\mu A/\mu m^2}$ & $\mathrm{\mu m}$ & $\mathrm{\Omega\,\mu m^2}$ & $\mathrm{\Omega\,\mu m}$ \\
  \hline
  I-a & 2.3(2) & 1.39(9) & 860(2) & 3797(3) \\
  I-b & 1.1(3) & 1.0(5) & 1655(5) & 7347(9) \\
  I-c & 0.4(3) & 0(1) & 1754(7) & 9488(2) \\
  II-d & 24.7(2) & 0.13(1) & 64(6) & 189(2) \\
  III-d & 9.4(1) & 0.46(1) & 173(4) & 385(4) \\
  \hline
\end{tabular}
\end{table}

\end{document}